\documentclass[twocolumn]{aastex701}

\usepackage{amsmath} 
\usepackage{enumerate}
\usepackage{natbib}
\usepackage{upgreek}

\usepackage[version=4]{mhchem}

\newcommand\aastex{AAS\TeX}

\newcommand\virga{\texttt{Virga}}

\defcitealias{moran_and_lodge_2025}{Moran \& Lodge et al. 2025}

\shorttitle{Fast Fluffy Fractals for {\virga} II}
\shortauthors{Lodge and Moran et al.}

\received{September 9, 2025}
\revised{November 7, 2025}
\accepted{November 30, 2025}

\begin{document}

\title{Template \aastex v7.0.1 Article with Examples\footnote{Footnotes can be added to titles}}

\title{Fractal Aggregate Aerosols in the Virga Cloud Code II: Exploring the Effects of Key Cloud Parameters in Warm Neptune, Hot Jupiter and Brown Dwarf Atmospheres \\ }

\author[0000-0002-9733-0617]{Matt G. Lodge}
\affiliation{School of Physics, HH Wills Physics Laboratory, Tyndall Avenue, Bristol, BS8 1TL, UK}
\email[show]{m.g.lodge@bristol.ac.uk}  
\correspondingauthor{Matt G. Lodge \& Sarah E. Moran}

\author[0000-0002-6721-3284]{Sarah E. Moran}
\altaffiliation{NHFP Sagan Fellow}
\affiliation{NASA Goddard Space Flight Center, 8800 Greenbelt Rd, Greenbelt, MD 20771, USA}
\email[show]{sarah.e.moran@nasa.gov}

\author[0000-0003-4328-3867]{Hannah R. Wakeford}
\affiliation{School of Physics, HH Wills Physics Laboratory, Tyndall Avenue, Bristol, BS8 1TL, UK}
\email{hannah.wakeford@bristol.ac.uk}

\author[0000-0003-4813-7922]{Zo{\"e} M. Leinhardt}
\affiliation{School of Physics, HH Wills Physics Laboratory, Tyndall Avenue, Bristol, BS8 1TL, UK}
\email{zoe.leinhardt@bristol.ac.uk}

\author[0000-0002-5251-2943]{Mark S. Marley}\affiliation{Department of Planetary Sciences and Lunar and Planetary Laboratory, University of Arizona, 1629 E University Blvd, Tucson, AZ 85721, USA}\email{marksmarley@arizona.edu}

\begin{abstract}

Aerosols and clouds are expected to be ubiquitous in exoplanet and brown dwarf atmospheres, where they can have a significant impact on transmission and emission spectra. The cloud code {\virga} is capable of quickly modeling cloud particle sizes as a function of altitude, and has recently been updated to include functionality for aggregates (ranging from very fluffy chains to compact fractals). We analyze the effect that these aggregates have on transmission spectra for typical warm Neptune and hot Jupiter environments, as well as their effect on emission spectra for an L-type brown dwarf, over the wavelength range 0.3 - 15 $\mu$m. We find significant, measurable differences in spectra when particle shape is changed (particularly the shortest wavelengths where particle morphology strongly affects the scattering slope). We provide some intuitive rules for how non-absorbing aggregates impact spectra: when particle sizes are small compared to the wavelength of light, the most elongated and chain-like particles have the highest opacities. When particles are large, the inverse is true (the most compact shapes have the highest opacities). We present an explanation for these effects in terms of the dynamics of how the particles form and move through the atmosphere, as well as in terms of fundamental optics theory. Given the significant impact that particle shape can have on spectra, we strongly encourage the community to include shape as a free parameter in future case studies, atmospheric models, and retrievals.

\end{abstract}



\section{Introduction}

Aerosols in the form of condensate clouds and photochemical hazes have critical, wide-ranging importance in astrophysical observations, and have been detected in every major body in the solar system. Recent observations from exoplanet and brown dwarf atmospheres \citep[e.g.,][]{grant2023jwst, biller2024,dyrek2024,hoch2025silicate} require suspended atmospheric particles to explain spectral features, where their wavelength-dependent opacity can affect the intensity of transmission and emission spectra \citep[e.g.,][]{morley2012, morley2013, sing2016, benneke2019, feinstein2023, miles2023, mak20243d, mak2025advancing}. When fitting models to data, aerosols are often either represented through parametrization (e.g., \texttt{POSEIDON}, \citealt{macdonald2017hd, macdonald2024poseidon, Mullens2024}; \texttt{petitRadTrans}, \citealt{Molliere2019,nasedkin2023atmospheric}; \texttt{Picaso}, \citealt{picaso,Mukherjee2023}; \texttt{Brewster}, \citealt{Burningham2017,2021MNRAS.506.1944B}), or by using microphysical models (e.g., \texttt{CARMA}, \citealt{Toon1988,Powell2018ApJ...860...18P}; \texttt{DRIFT/StaticWeather}, \citealt{Helling2017A&A...603A.123H}; \texttt{ARCiS}, \citealt{ormel2019arcis}; \texttt{mini-cloud}, \citealt{lee2025three}). For the most recent review of these models, see \citet{gao2021}. Because of the computational requirements of the more complex models, it is often intractable to explore a wide parameter space when attempting to match observational data. Simpler 1D models (e.g., {\virga}, \citealt{Batalha2020, batalha2025, Rooney2022}, \texttt{ExoLyn}; \citealt{huang2025exolyn, huang2024exolyn}) enable the community to predict the effects of aerosols on spectra, while retaining fast computation times, to allow the full range of diverse chemistry and particle sizes/distributions within aerosols to be explored. 

The majority of models consider all particles to be spherical as a first-order approximation of their shape, but several studies have demonstrated that the shape of the particles can have a significant impact on the optical and dynamical properties, and therefore have an impact on atmospheric spectra \citep[e.g.,][]{ohno2018microphysical, ohno2020clouds,Samra2020,2024MNRAS.527.4955C, vahidinia2024aggregate, lodge2024aerosols, lodge2024manta, hamil}. 

The cloud code {\virga}, based on \texttt{eddysed} \citep{ackerman2001}, has been recently updated to include fractal aggregates and thus the effects of particle shape beyond spheres \citepalias{moran_and_lodge_2025}. As the first paper in the series, \citetalias{moran_and_lodge_2025} laid out the model framework and updates needed in {\virga} to account for the dynamical and optical behavior of fractal aggregates. However, that paper demonstrated only a limited range of potential uses of the new code functionality.  Here, in this second paper of the series, we widen the scope of the code application. Rather than only changing the fractal shapes as in \citetalias{moran_and_lodge_2025}, we also explore how {\virga}'s primary variables -- the eddy diffusion coefficient, $K_{\rm{zz}}$, and the sedimentation efficiency, $f_{\rm{sed}}$ -- affect the fractal aggregate cloud structures produced by {\virga} for a range of frequently studied substellar environments. To gain insight into how aggregates might influence observations of these worlds, we explore aggregate cloud particles made of both KCl and amorphous \ce{Mg2SiO4} in the atmospheres of warm Neptune, hot Jupiter, and brown dwarf environments.

In section~\ref{sec:Methods} we briefly summarize the new methodology used for aggregates in {\virga}, and also make some basic predictions on how they might impact spectra. Additionally, we describe the \texttt{PICASO} model atmospheres used to generate synthetic transmission and emission spectra in each case study. In section~\ref{sec:Results} we discuss the key effects of considering aggregates, in terms of the particle sizes/number densities and overall opacity. We provide example spectra for two key tunable parameters within {\virga} and explain the observed effects. In section~\ref{sec:Conclusions}, we summarize the key `rules of thumb' for the behaviour of aggregates within {\virga}, and make recommendations going forward.

\section{Methodology} \label{sec:Methods}
In this section we outline the \texttt{Virga} model (\S \ref{subsec:virga-model}), detail the aggregate properties (\S \ref{subsec:properties}), optics (\S \ref{subsec:optics}), and dynamics (\S \ref{subsec:dynamics}), and outline the configuration of our atmosphere case studies (\S \ref{subsec:case-studies}). 

\subsection{\texttt{Virga} Model}\label{subsec:virga-model}

{\virga} \citep{Batalha2020,Rooney2022,batalha2025} is an open-source python-based cloud code, based on the \texttt{EddySed} model \citep{ackerman2001, GaoMarleyAckerman2018}. This 1D model stratifies pressure layers in the atmosphere, and at each altitude balances upwards transport (e.g., from convection, eddies, or otherwise bulk upwards movement of material), and downwards sedimentation (due to gravity) as governed by:
\begin{equation} \label{Eq:eddysed}
    -K_{zz}\frac{\delta q_t}{\delta z} - f_\mathrm{sed} w_* q_c=0.
\end{equation}
Here $K_{zz}$ is the eddy-diffusion coefficient, which parametrizes strength of the upwards motion and can be calculated from Global Circulation Models (GCMs) or estimated by the user (and can be fixed or a function of altitude). $f_\mathrm{sed}$ is the sedimentation efficiency (the ratio of the sedimentation velocity to the convective velocity).  $f_\mathrm{sed}$ therefore effectively becomes a variable that controls particle size (i.e., small $f_\mathrm{sed}$ causes smaller particles spread over a wider pressure range). $q_c$ is the condensate mass mixing ratio, $q_t$ is the total (vapor + condensate) mass mixing ratio, $z$ is the altitude, and $w_*$ is the mean upward velocity in each pressure layer. {\virga} is well-tested for spherical particles (for a summary of successful studies, see \citealt{batalha2025}), and now includes functionality for aggregate particles (\citetalias{moran_and_lodge_2025}). We explore the interplay of fractal aggregates with varying $f_\mathrm{sed}$ and  $K_{zz}$ in this work. 

\subsection{Aggregate Properties}\label{subsec:properties}

\begin{figure}
    \includegraphics[width=\columnwidth]{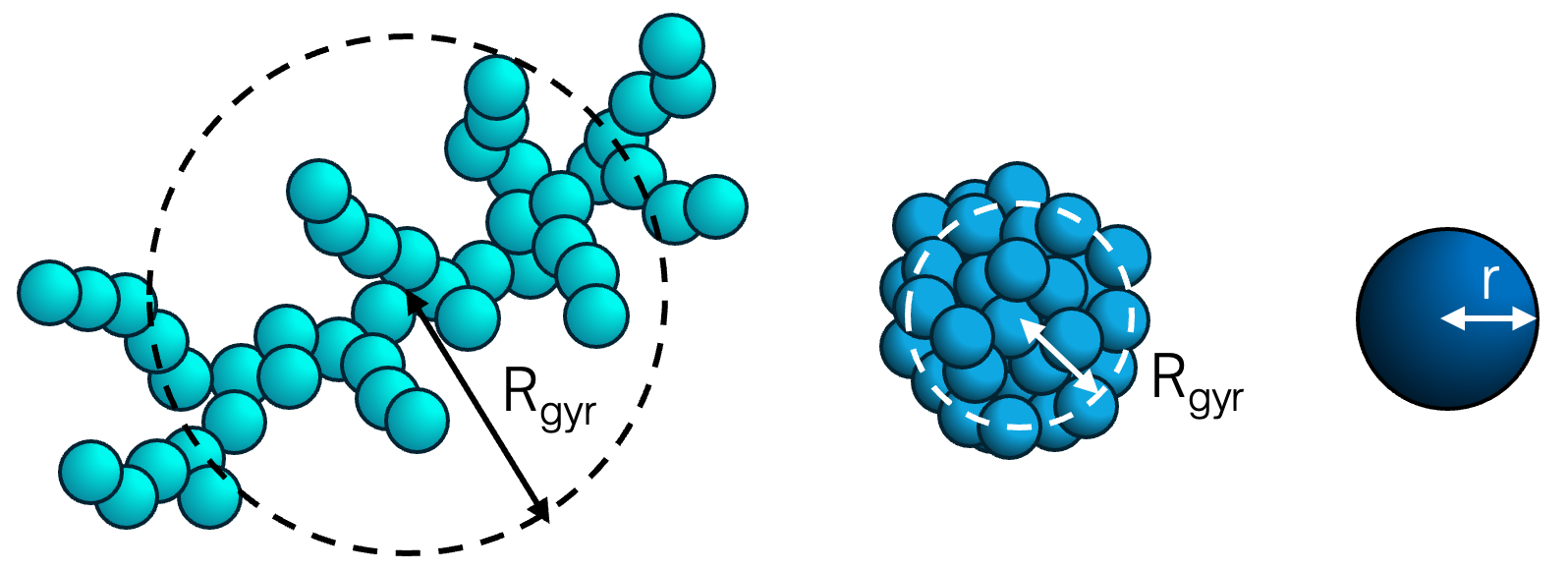}
    \caption{Three possible shapes for aerosol particles (from left to right, fractal dimensions $D_{\rm{f}}$ of 1.6, 2.8, and a sphere). In this illustration each shape has the same mass, and therefore also the same compact radius $r$ (the radius of a sphere made from the same volume as each shape). However, the radius of gyration $R_{\rm{gyr}}$ of the aggregates (used to calculate the optical and dynamical properties) can differ, and is largest for shapes with the lowest fractal dimensions.}
    \label{fig:fractals}
\end{figure}

Aggregates approximate well to a fractal scaling relation, where they are modeled to be comprised of $N_{\rm{mon}}$ smaller monomers of radius $r_{\rm{mon}}$, joined together in a shape primarily defined by the fractal dimension $D_{\rm{f}}$:
\begin{equation} \label{eq:fractals}
    N_{\rm{mon}} = k_0 \left( \frac{R_{\rm{gyr}}}{r_{\rm{mon}}} \right)^{D_{\rm{f}}},
\end{equation}
where $k_0$ is the fractal prefactor and $R_{\rm{gyr}}$ is the radius of gyration. In {\virga} either $N_{\rm{mon}}$ or $r_{\rm{mon}}$ can be fixed and in \citetalias{moran_and_lodge_2025} we explore the effects of both of these model options in depth. In reality, aerosols are likely to be heterogeneous (as they are on Earth), and not built from monomers that maintain identical sizes across the many orders of magnitude spanned by aerosol aggregate sizes (fixed $r_{\rm{mon}}$). Nor are they likely to be composed of a single aggregate shape that scales perfectly with size (fixed $N_{\rm{mon}}$). The truth is likely somewhere in-between these two models, and exploring each representation separately can provide useful insight. Here we choose to fix $N_{\rm{mon}}=1000$ to avoid computational instabilities that can be introduced by fixing $r_{\rm{mon}}$ \citepalias{moran_and_lodge_2025}. In this model, $N_{\rm{mon}}$ can still be smaller than 1000 to allow very small particles to form (e.g. made of only a few monomers of the smallest possible physical size). Where $N_{\rm{mon}} \geq 100$, we use the linearly interpolated expression for $k_0$ proposed by \citet{Tazaki2021}, which captures the behavior of aggregates formed from cluster-cluster aggregation to linear-chains, valid for 1~$\le~D_{\rm{f}}~\le$~3 (to within 5 \% accuracy):
\begin{equation} \label{eq:k_0}
    k_0 = 0.716(1 - D_{\rm{f}}) + \sqrt{3}.
\end{equation}
For cases where $N_{\rm{mon}} < 100$, we create a linear fit such that $k_0$ smoothly transitions to $k_0=1$ at $N_{\rm{mon}}=1$ for any fractal dimension (as in \citetalias{moran_and_lodge_2025}):
\begin{equation}
    k_0 = \frac{1.448-0.716D_{\rm{f}}}{99}\left(N_{\rm{mon}} - 1 \right) +1.
\end{equation}
The radius of gyration is defined as the radial distance to a point which would have the same moment of inertia as the body's actual distribution of mass, if the total mass of the body were concentrated at that point. $R_{\rm{gyr}}$ scales with particle size and shape (see Fig.~\ref{fig:fractals}) in a way that makes it an important characteristic radius for aggregates (see \citetalias{moran_and_lodge_2025} for full details on its use in both the optics and dynamics within {\virga}). We highlight that some studies compare aggregates using $R_{\rm{gyr}}$, but when comparing between shapes in this paper, we use the compact radius $r$ (the radius of a sphere of equivalent volume to the aggregate). This is an important distinction, because shapes with the same compact radius will also have the exact same mass -- this allows us to directly compare the optics and dynamics of two aggregates of varying $D_{\rm{f}}$ where the only difference is the particle shape. Using this fractal model, shapes with $D_{\rm{f}}$ values closer to 1 are more elongated and fluffy, whereas $D_{\rm{f}}$ approaching 3 are more compact, as shown in Figure~\ref{fig:fractals}.

\subsection{Aggregates: Dynamics}\label{subsec:dynamics}

\begin{figure*}
    \includegraphics[width=\textwidth]{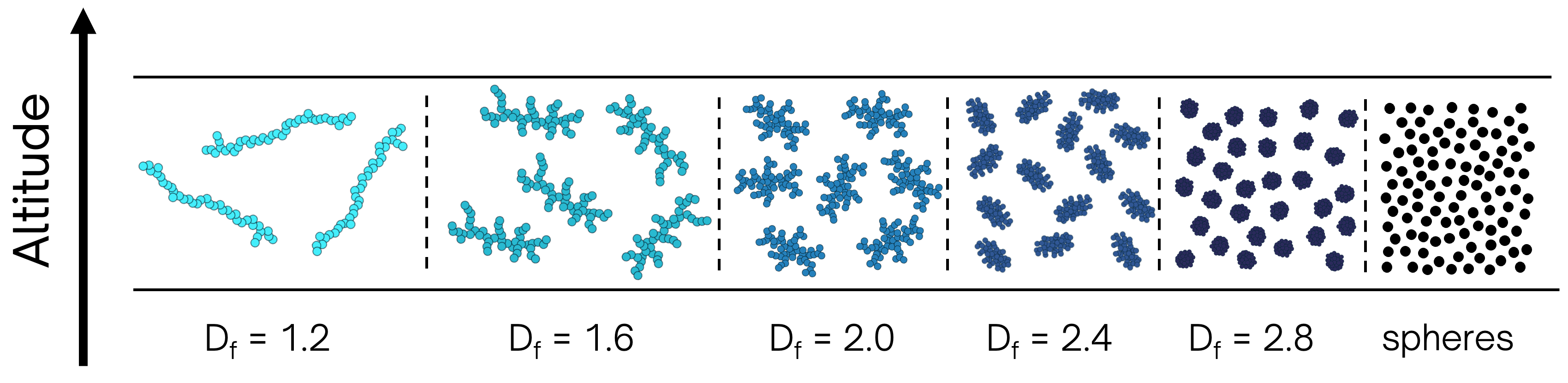}
    \caption{Comparison of the number densities and sizes of particles formed within a single pressure layer for different condensate particle shapes. In the actual {\virga} model, distributions follow a lognormal size distribution, but here we only show the mean particle size in each case for clarity in showing how the sizes of particles within the distributions compare.}
    \label{fig:big_fractals}
\end{figure*}

Regardless of the aggregate shape that is modeled, the same total mass of vapor condenses out into solid particles in each pressure layer. However, the terminal velocity for an aggregate is always smaller than for spheres of the same mass, because of how their shape changes the dynamical interaction with the surrounding medium (\citetalias{moran_and_lodge_2025}). Therefore a key difference between aggregates and spheres is that aggregate particles are required to have more mass for their terminal velocities to equal the upwards velocity in a given pressure layer ($v_{\rm{fall}}=w_*$). Aggregates with the lowest fractal dimensions form into the largest particles, with much fewer number densities (because the same total amount of mass is available to condense in a given layer). This is demonstrated visually in Figure~\ref{fig:big_fractals}. The distributions of aggregates are therefore quite different and have very different dynamical and optical properties to their spherical counterparts, which in turn has measurable consequences on transmission and emission spectra.

\begin{figure}
    \includegraphics[width=\columnwidth]{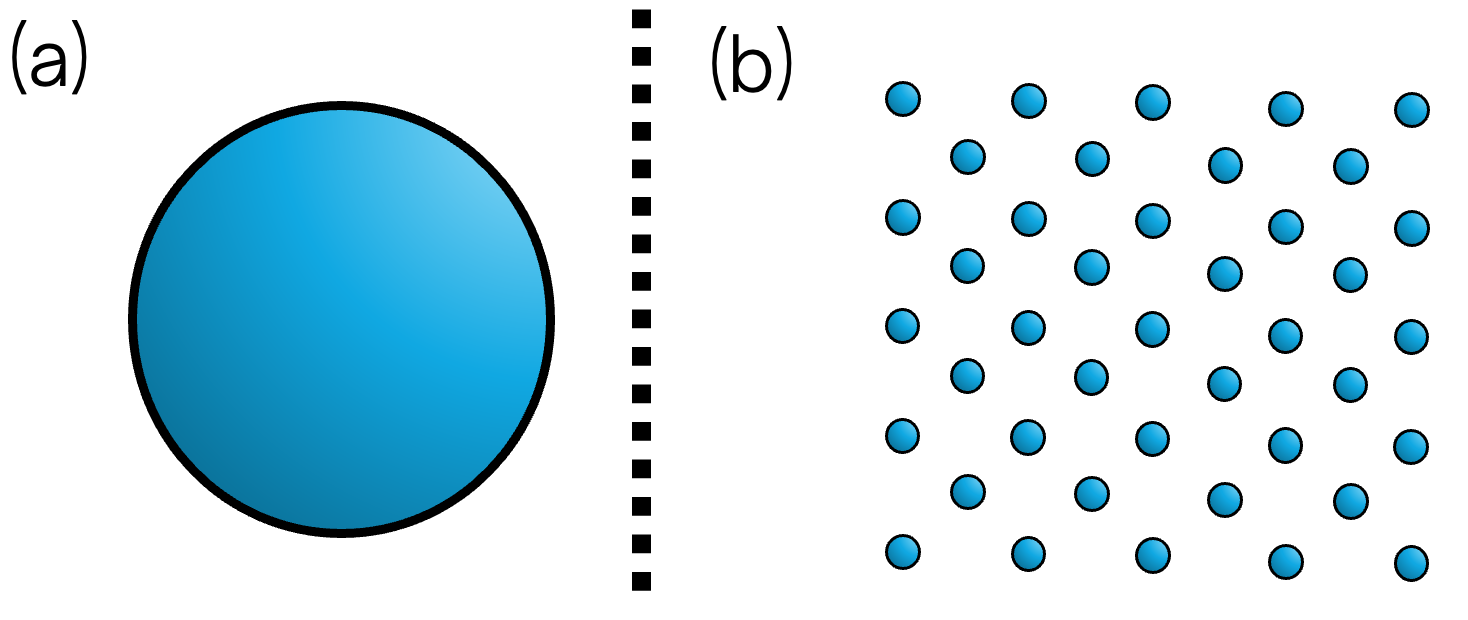}
    \caption{Diagram describing a toy model in which we consider the difference in opacity of a single sphere of radius $R$, versus the same mass but distributed into a group of much smaller spheres of radius $r$.}
    \label{fig:toy_model}
\end{figure}

\subsection{Aggregates: Optics}\label{subsec:optics}

\begin{figure*} 
    \includegraphics[width=\textwidth]{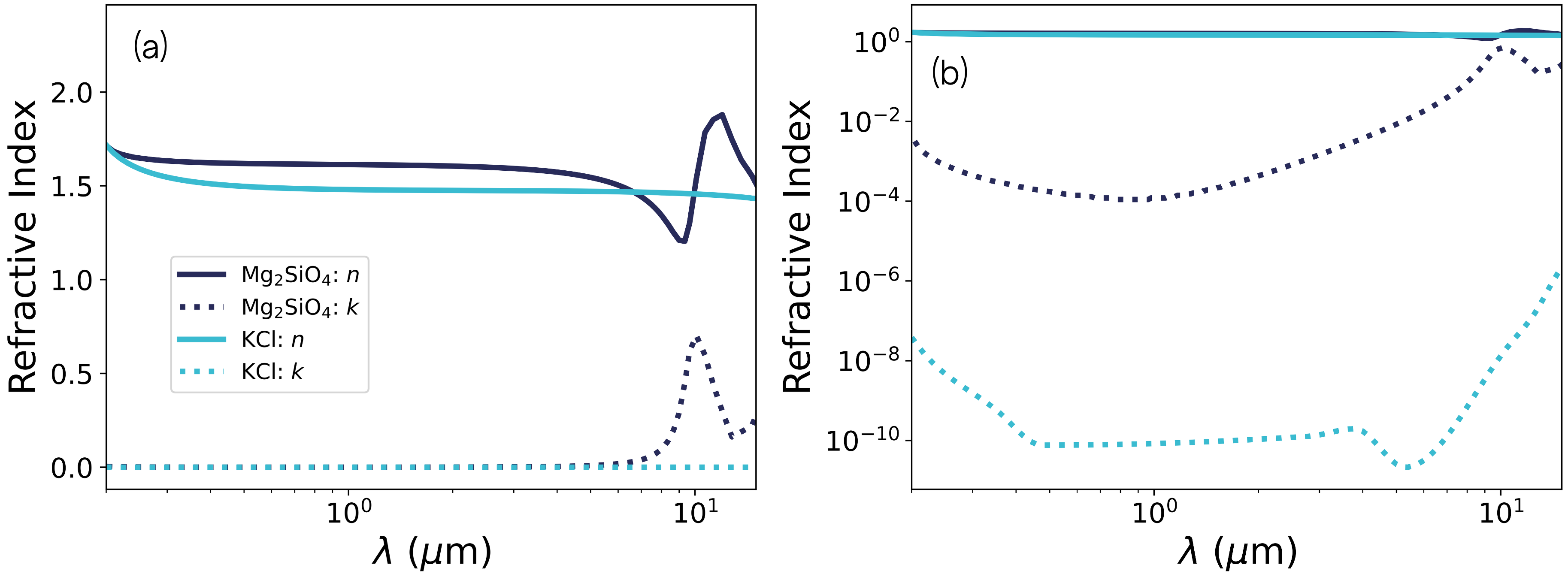}
    \caption{Refractive indices of KCl (light blue) and Mg$_2$SiO$_4$ (dark blue) as a function of wavelength, plotted on (a) linear and (b) logarithmic scales. The imaginary component $k$ (dotted lines) determines how absorptive each material is at a particular wavelength.}
    \label{fig:refractive_index}
\end{figure*}

\begin{figure*}
    \includegraphics[width=\textwidth]{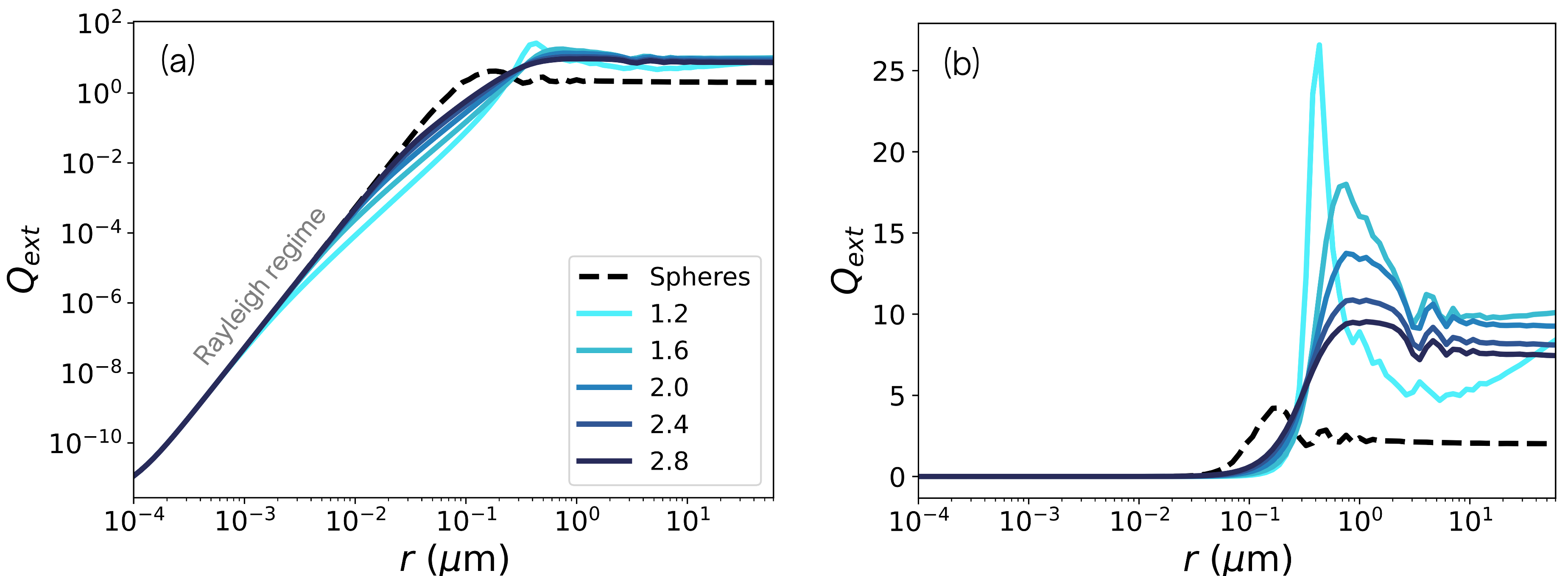}
    \caption{Extinction efficiency as a function of compact radius $r$ for KCl particles of a variety of different shapes (with $\lambda=0.3~\mu$m), plotted on both a logarithmic (a) and linear (b) scale. The Rayleigh regime (where $r \ll \lambda$ and therefore the approximation $Q_{\rm{ext}}\propto r^4$ can be used) is shown in (a).}
    \label{fig:Q_ext_vs_r}
\end{figure*}

{\virga} uses Modified Mean Field (MMF) theory (via calling the package \texttt{optool}; \citealt{dominik2021optool}) to calculate the optical properties of aggregates \citep{Tazaki_Tanaka_2018, berry1986optics, botet1997mean, dominik2021optool}. The key metric dictating overall absorbing and scattering due to a particle is the extinction efficiency $Q_{\rm{ext}}$, which is a function of refractive index, wavelength, and particle radius. Figure~\ref{fig:refractive_index} shows how the refractive index of KCl \citep{querry1987} and amorphous Mg$_2$SiO$_4$ \citep{jaeger2003} vary as a function of wavelength. KCl has such low values of $k$ that the opacity is purely from scattering (and not absorption) at these wavelengths. Figure~\ref{fig:Q_ext_vs_r} shows how $Q_{\rm{ext}}$ varies with compact radius $r$ for KCl at the shortest wavelength considered in this study ($\lambda=0.3~\mu$m). Figure~\ref{fig:Q_ext_vs_r}b shows that the optical properties of aggregates can differ significantly from spheres, even when particles have the same amount of mass (represented by them having the same compact radius). MMF is designed such that the optical properties of aggregates and spheres are approximately equal for particles that are small compared to the wavelength (often referred to as the long-wavelength limit -- this regime can be seen on the far-left of Figure~\ref{fig:Q_ext_vs_r}a). This approximation holds for the moderate refractive index materials in this paper, but may be less valid for high refractive index materials such as TiO$_2$ or crystalline, rather than amorphous, \ce{Mg2SiO4}. Integrating {\virga} and MMF with a more-advanced long-wavelength limit model \citep{lodge2024manta} will be the subject of a future study.

When particles are very large compared to the wavelength, the memory requirements of \texttt{optool} can become very high. However, the extinction efficiencies are constant in this regime (the geometric regime, where size parameter $x=\frac{2 \pi r}{\lambda} \gg 1$) as shown on the right of Figure~\ref{fig:Q_ext_vs_r}a and b. Therefore, for the extreme cases where the particles are very large (size parameter $x>2000$), we extrapolate $Q_{\rm{ext}}$ as a constant value to save computation time and to allow calculations for the largest particles that would otherwise be outside the bounds of available computational memory. 

To gain some intuition into how the opacity of each of these aggregate models will compare, we begin with analysis of a simplified toy model. Consider an example where a single large sphere of radius $R$ is broken into $N$ smaller spheres of radius $r$ (Fig.~\ref{fig:toy_model}). Here the total mass and volume of the spheres in each case is conserved, such that: 
\begin{equation}
    \frac{4}{3} \pi R^3 = N \frac{4}{3} \pi r^3,
\end{equation}
And therefore for a fixed $R$: 
\begin{equation} \label{eq:N_spheres}
    N \propto r^{-3}.
\end{equation}
The total mass-weighted opacity of the distribution of small spheres is calculated using:
\begin{equation} \label{eq:kappa}
    \kappa = \frac{N Q_{\rm{ext}} \pi r^2}{M}
\end{equation}
where $N$ is the number of spheres, $Q_{\rm{ext}}$ is the extinction efficiency (which is a function of sphere radius $r$) and $M$ is the total mass (which is constant in the two cases shown above). We can calculate opacity for both of the cases shown in Figure~\ref{fig:toy_model}a and Figure~\ref{fig:toy_model}b in two different size regimes: the Rayleigh regime and the geometric regime, described in detail in the following subsections.

\subsubsection{Small particles (Rayleigh regime)} \label{sec:rules_small_particles}

Some chemical species of interest in exoplanet/brown dwarf atmospheres are essentially non-absorbing at near ultraviolet and visible wavelengths (e.g. KCl, and Mg$_2$SiO$_4$ at some particle sizes/wavelengths). The exact proportion of radiation that is absorbed/scattered depends on particle size, wavelength and refractive index, but here we consider all small KCl particles to be purely scattering, due to their extremely low imaginary refractive index $k$ (see Figure~\ref{fig:refractive_index}). Mg$_2$SiO$_4$ particles in this context are mostly non-absorbing, but the smallest particles ($r<0.01~\mu$m) may have a significant absorbing component, and the simplified analysis presented below would not be applicable to these. For non-absorbing particles that are also very small in size (i.e., radius much smaller than the wavelength of radiation, $r \ll \lambda$), the extinction efficiency is given by:
\begin{equation} \label{eq:Q_ext_rayleigh}
    Q_{\rm{ext}} \approx Q_{\rm{sca}} \approx \frac{8}{3} \left(\frac{2 \pi r}{\lambda} \right)^4   \left(\frac{n^2-1}{n^2+2} \right)^2,
\end{equation}
where $n$ is the real refractive index of the material. For a fixed wavelength, we therefore have $Q_{\rm{ext}}\propto r^4$ (as shown on the left side of Fig.~\ref{fig:Q_ext_vs_r}a). Substituting Eq.~\ref{eq:N_spheres} and \ref{eq:Q_ext_rayleigh} into Eq.~\ref{eq:kappa}:
\begin{align}
    \kappa &\propto r^{-3} r^4 r^2 \\
    \kappa &\propto r^{3}. \label{eq:opacity_rayleigh}
\end{align}
Therefore in this regime, the opacity is larger when particles form into a single particle with a large radius (as in Fig.~\ref{fig:toy_model}a), compared to when the same total mass is split into lots of smaller particles (as in Fig.~\ref{fig:toy_model}b). Considering realistic aggregate shapes, we earlier demonstrated that the elongated fluffy fractals always form fewer, but larger, particles (Fig.~\ref{fig:big_fractals}). Therefore in this regime, we might also expect the fluffier particles (on the left on Fig.~\ref{fig:big_fractals}) to have a higher total opacity than the more numerous (but smaller) compact aggregates (on the right on Fig.~\ref{fig:big_fractals}). While the leap from our toy model (spheres) to aggregates might initially seem like a large step, the optical properties for spheres and aggregates in this regime are identical when using MMF, so Eq.~\ref{eq:Q_ext_rayleigh} is true regardless of particle shape, and the assumptions in the toy model are valid for aggregates in this regime as well.

\subsubsection{Large particles (Geometric regime)} \label{sec:rules_large_particles}

For particles that are much larger than the wavelength, $Q_{\rm{ext}} \approx \rm{constant}$ for a particular shape, as seen on the right of Figure~\ref{fig:Q_ext_vs_r}. The one exception is for $D_{\rm{f}}=1.2$, but we note that this is an extreme shape which is pushing the optical modeling (as well as the dynamical modeling; see \citetalias{moran_and_lodge_2025}) to the limit, and as such the results should be interpreted with caution. By setting $Q_{\rm{ext}}$ as constant and substituting Eq.~\ref{eq:N_spheres} in Eq.~\ref{eq:kappa}:
\begin{align}
    \kappa &\propto r^{-3} r^2 \\
    \kappa &\propto \frac{1}{r} \label{eq:opacity_large_particles}
\end{align}
This is the opposite relationship to the one found for small particles (Eq.~\ref{eq:opacity_rayleigh}). For large particles, the larger fluffier aggregates with low fractal dimensions (Fig.~\ref{fig:big_fractals}, left) are less opaque than the smaller more compact aggregates (Fig.~\ref{fig:big_fractals}, right). 

We therefore have two general rules for how aggregates will affect spectra (Eq.~\ref{eq:opacity_rayleigh} and Eq.~\ref{eq:opacity_large_particles}), depending on the size of particles that form. It is also worth noting that the transition between these two regimes is not necessarily smooth, because where the particle size and radius are roughly equal there are optical resonances, as shown in Figure~\ref{fig:Q_ext_vs_r}. The effects that the transition region has on spectra will be discussed in section~\ref{sec:Results}.

\subsection{Model Atmosphere Case Studies}\label{subsec:case-studies}

To demonstrate the effects of including aggregate cloud particles in a variety of environments, we explore three case studies with varying atmospheric conditions. All of these atmospheres use the radiative-convective thermochemical equilibrium climate and radiative transfer code \texttt{PICASO} \citep{picaso,Mukherjee2023}. Our case studies include a range of equilibrium (or effective, in the case of the brown dwarf) temperatures and masses: a warm Neptune, a hot Jupiter, and a brown dwarf. The parameters of each object are chosen to be representative of a ``typical'' object in each class rather than that of a specific observational target; for both the hot Jupiter and brown dwarf, we use the in-built temperature-pressure and chemical profiles included in the default release of \texttt{PICASO 3.0}. We describe the set-up and assumptions to each model atmosphere in the following subsections.

\subsubsection{Warm Neptune}

For the warm Neptune, we create a new generic atmospheric model using \texttt{PICASO 3.0} \citep{Mukherjee2023}, based in part on the planetary parameters of the well-studied HAT-P-26~b and HAT-P-11~b. Therefore, the climate module is run for a 0.5 R$_{\rm{Jup}}$, 0.05 M$_{\rm{Jup}}$ object with an intrinsic temperature of 200 K and an equilibrium temperature of 800 K, with heat
redistribution factor 0.6 to account for near full day-night heat circulation. The stellar parameters are that of a 5700~K T$_{\rm{eff}}$, log($g/\rm{cm}~\rm{s}^{-2}$)~=~4.5, 1~R$_\odot$ star with solar metallicity. The planet is assumed to be at 0.16 au.

\begin{figure}
    \includegraphics[width=\columnwidth]{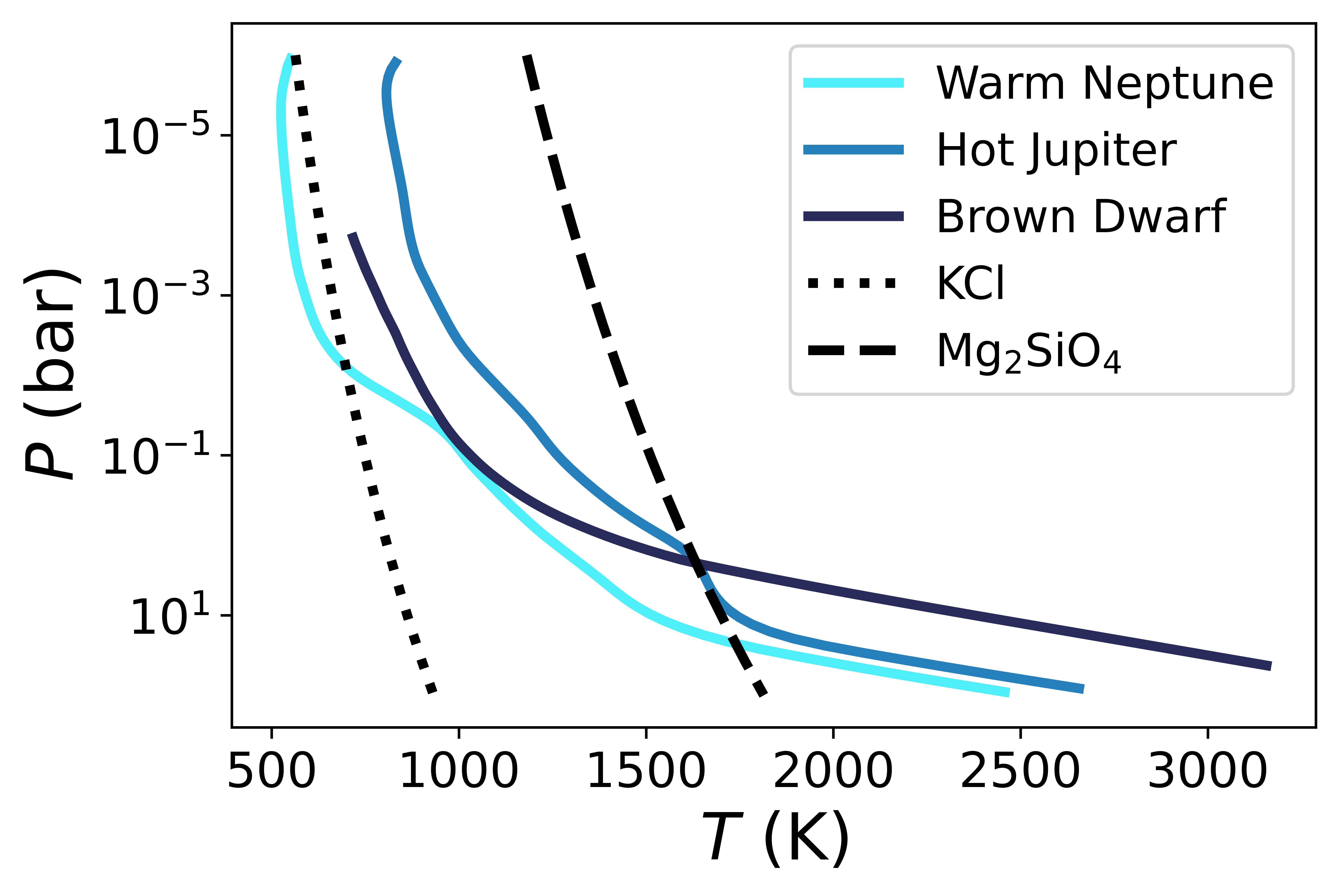}
    \caption{The pressure-temperature profiles for the warm Neptune, hot Jupiter and brown dwarf case studies, as well as condensation curves for the two key condensing species in this paper (KCl and \ce{Mg2SiO4}). The intersection of the p-T profile and condensation curve represents the point above which clouds of a given species will form.}
    \label{fig:pt_profiles}
\end{figure}

We initiate the climate run using the correlated-k coefficients of \citet{lupu2021} at 100$\times$ solar metallicity and a C/O ratio of 0.458, following the \citet{Lodders2009} solar value. Pre-weighted opacities in \citet{lupu2021} include \ce{C2H2}, \ce{C2H4}, \ce{C2H6}, \ce{CH4}, CO, \ce{CO2}, CrH, Fe, FeH, \ce{H2}, \ce{H3+}, \ce{H2O}, \ce{H2S}, HCN, LiCl, LiF, LiH, MgH, \ce{N2}, \ce{NH3}, OCS, \ce{PH3}, SiO, TiO, VO, Li, Na, K, Rb, and Cs. The temperature-pressure profile (Fig. \ref{fig:pt_profiles}) is parameterized with initial conditions following a \citet{Guillot2010} profile with 91 pressure levels bounded by 100 to 10$^{-6}$ bar. The uppermost level of the convective zone guess starts at 85, with one convective zone. We iterate from these initial conditions until convergence. The resulting temperature profile ranges from 557 K at the top of the atmosphere to 2457 K at 90 bar.

Given the temperature profile of the planet, along with the 100$\times$ solar metallicity atmosphere, we use KCl as our primary cloud condensate of interest with a mass mixing ratio of $8.635\times 10^{-6}$ (the default value in \virga). For KCl refractive indices, we use the data of \citet{querry1987}, shown in Figure~\ref{fig:refractive_index}. KCl is expected to be the dominant cloud species at this temperature from an observational perspective \citep{Gao2020}, given its high nucleation rate and elevated abundance in the atmosphere \citep{gaobenneke2018}. KCl has also been previously used in aggregate studies for similar objects, like the sub-Neptune GJ~1214~b (\citealt{ohno2020clouds}; \citetalias{moran_and_lodge_2025}). 

\subsubsection{Hot Jupiter}

For our hot Jupiter, we use the existing profile contained within \texttt{just\_do\_it} in all versions of {\virga}. This profile consists of an atmosphere that was created with \texttt{PICASO 1.0} \citep{picaso}, with pressure bounds from 1.1$\times$10$^{-6}$ to 90 bar and temperatures from 806 to 2656 K. The atmospheric chemistry is for 1$\times$ solar metallicity and a solar C/O ratio, and includes H$_2$, H, He, \ce{H2O}, \ce{CH4}, CO, \ce{NH3}, \ce{N2}, \ce{PH3}, \ce{H2S}, \ce{TiO}, \ce{VO}, \ce{Fe}, \ce{FeH}, \ce{CrH}, \ce{Na}, \ce{K}, \ce{Rb}, \ce{Cs}, \ce{CO2}, \ce{HCN}, \ce{C2H2}, \ce{C2H4}, \ce{C2H6}, \ce{SiO}, \ce{MgH}, \ce{OCS}, \ce{Li}, \ce{LiOH}, \ce{LiH}, \ce{LiCl}, and \ce{LiF} \citep[with opacities from][]{lupu2021}. We set the planetary radius and mass at 1~R$_{\rm{Jup}}$ and 1~M$_{\rm{Jup}}$. For the stellar parameters, we use identical values to the warm Neptune case: $T_{\rm{eff}}=5700$~K, log($g/\rm{cm}~\rm{s}^{-2}$)~=~4.5, solar metallicity, $R=1$~R$_\odot$. 

For the condensible species, we choose \ce{Mg2SiO4} with a mass mixing ratio of $3.796\times 10^{-3}$. As with KCl in our warm Neptune, \ce{Mg2SiO4} is expected to be both highly abundant and easy to condense under hot Jupiter conditions due to its high heterogeneous nucleation rate \citep{Lee2016,Gao2020}. The exact species of silicate --- i.e., \ce{SiO2}, \ce{Mg2SiO4}, \ce{MgSiO3}, SiO -- will depend on the specific conditions of the atmosphere \citep{Lee2016,Calamari2024,moran2024neglected}. Crystalline \ce{Mg2SiO4} has multiple peaks observable in the JWST MIRI/LRS wavelength range, while other forms only have one peak \citep[e.g.,][]{wakeford2015,kitzmannheng2018,LunaMorley2021,Mullens2024}. For simplicity, we use the amorphous form of \ce{Mg2SiO4} refractive indices, as shown in Figure~\ref{fig:refractive_index}, using the data of \citet{jaeger2003}. We use this amorphous \ce{Mg2SiO4} as our representative silicate with which to examine fractal aggregate particle behavior in hot Jupiter atmospheres though we note that other silicate forms may have their own distinct behavior when applied to fractal aggregates.

\subsubsection{Brown Dwarf}

Our brown dwarf atmosphere comes from the \texttt{Sonora} series of models. In this case, our base atmosphere is taken from the cloudless, equilibrium \texttt{Sonora Bobcat} series \citep{marley2021}. The atmosphere is that of a 1700 K $T_{\rm{eff}}$, log($g/~\rm{cm}~\rm{s}^{-2}$)~=~5 L-dwarf with solar metallicity and C/O ratio. This atmosphere includes chemical contributions from \ce{H2}, H, H-, H+,  \ce{H2}-, \ce{H2}+, \ce{H3}+, He, \ce{H2O}, \ce{CH4}, CO, \ce{NH3}, \ce{N2}, \ce{PH3}, \ce{H2S}, TiO, VO, Fe, FeH, CrH, Na, K, Rb, Cs, \ce{CO2}, HCN, \ce{C2H2}, \ce{C2H4}, \ce{C2H6}, SiO, MgH, OCS, Li, LiOH, LiH, LiCl, graphite, Al, and CaH. The atmosphere spans 91 pressure levels from 1.7$\times$10$^{-4}$ to 45 bar, with temperatures from 715 to 3193 K. 

We again choose \ce{Mg2SiO4} for our cloud species, as this temperature-pressure profile is very similar to our hot Jupiter, and silicates are the expected dominant clouds in L-dwarfs \citep[e.g.,][]{Cushing2006,Stephens2009,Suarez2022} as well as directly imaged planets of similar size and temperatures \citep[][]{hoch2025silicate}. As with our hot Jupiter, other silicates such as \ce{MgSiO3}, \ce{SiO2} \citep[e.g.,][]{2021MNRAS.506.1944B}, or SiO \citep[e.g.,][]{Molliere2025} are expected and observed depending on the brown dwarf's exact composition and conditions, but we focus here just on \ce{Mg2SiO4} for the same reasons as in our hot Jupiter case study.

\subsubsection{Transmission and Emission Spectra}

Once we have set the temperature-pressure and chemical conditions of substellar object as described in the preceding subsections, we generate model transmission (in the case of the warm Neptune and hot Jupiter) and emission (in the case of the L-dwarf) spectra. We use the radiative transfer capability of \texttt{PICASO} to account for the atmospheric opacity as a result of the temperature-pressure conditions and chemical abundances in each atmosphere. 

For the warm Neptune and hot Jupiter, we use the V3 opacity database available from Zenodo \citep{zenodo_opacities_v3}. This database covers wavelengths from 0.3 to 15 $\mu$m, and is resampled at R=15,000 from a line-by-line calculation performed at R$\sim$1e6. The database contains contributions from \ce{C2H2} \citep{hitran2012}, \ce{CH4} \citep{yurchenko13vibrational,yurchenko_2014}, CO \citep{HITEMP2010,HITRAN2016,li15rovibrational}, \ce{CO2} \citep{HUANG2014reliable}, \ce{H2-H2} \citep{Saumon12,Lenzuni1991h2h2}, \ce{H2O} \citep{Polyansky2018H2O}, K and Na \citep{Ryabchikova2015,Allard2007AA, Allard2007EPJD,Allard2016, Allard2019}, \ce{H2S} \citep{azzam16exomol}, OCS \citep{HITRAN2016}, and \ce{SO2} \citep{underwood2016exomol}. Our model atmospheric abundances contain more than these species. However, within the radiative transfer calculation, the additional molecules only contribute to the atmospheric mean molecular weight and thus the atmospheric scale height and resulting amplitude of spectral features, rather than as additional sources of absorption or scattering. The molecules included in the opacity database are the most dominant expected absorbing molecules for hot Jupiters and warm Neptunes from the optical to mid-infrared, and thus our model spectra should capture the main observational features expected in these atmospheres due to gas phase species. The resulting model transmission spectra are then rebinned to a resolution of R=300, consistent with median resolution of JWST data across the NIRISS, NIRSpec, and MIRI instrument modes.

For the brown dwarf spectra, the molecules TiO, VO, and metal hydrides such as FeH can strongly impact the emission at shorter (i.e., $<$1 $\mu$m) wavelengths for L-dwarfs \citep[e.g.,][]{kirkpatrick1999,marley2021}. Therefore we explored using the V2 opacity database on Zenodo \citep{zenodo_opacities_v2}, which contains these species. However, we found that only the clear spectra meaningfully differed when using this more extensive set of opacities, while all our cloudy models were not affected by the choice of opacity database. Therefore, we also use the V3 database described above 
for our brown dwarf spectra shown in this work. The resulting model emission spectra are then rebinned to a resolution of R=3000. As with the hot Jupiter and warm Neptune models, additional species included in the atmospheric chemistry (e.g., \ce{NH3}, \ce{PH3}, the metal hydrides, TiO and VO and the others listed in Section 2.5.3) affect only the scale height and molecular weight aspects of the radiative transfer calculation, rather than any additional opacity.

\section{Results} \label{sec:Results}

Here, we demonstrate how varying both the upward mixing and downward sedimentation in tandem with particle shape ultimately set the cloud layers, and thus the emergent spectra, of each of our model atmospheres. We find broadly that, as with spherical particles, increased $K_{\rm{zz}}$ results in larger particles overall, but that differing aggregate shapes enhance these particle size differences. Changes in $f_{\rm{sed}}$ similarly alter the particle size distributions, with larger values compressing the thickness of the cloud layer, but differing aggregate shapes more uniformly change within the bounds of changing sedimentation efficiency. In part, this uniform change with $f_{\rm{sed}}$ is by design, as the sedimentation efficiency is a non-physical quantity that parametrizes over detailed particle nucleation and coagulation in order to give a direct scaling of particle size with atmospheric pressure. Our key result is that aggregates can have predictable differences in opacity compared to spheres (either an increase or decrease), depending on a combination of fractal shape and particle size, which is determined by $K_{\rm{zz}}$, and $f_{\rm{sed}}$.

\begin{figure*}
    \includegraphics[width=\textwidth]{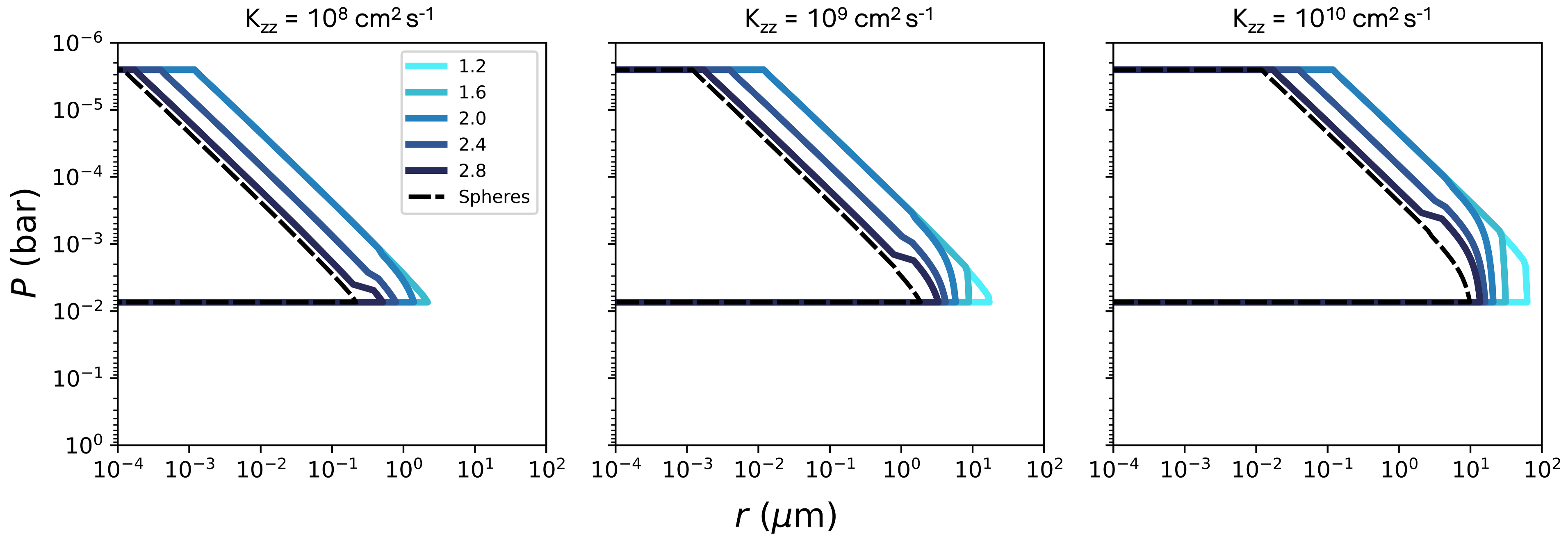}
    \caption{Mean sizes of KCl particles (in terms of compact radius) in cloud layers of a warm Neptune atmosphere, with $K_{\rm{zz}}=10^8$, $10^9$ and $10^{10}$ cm$^2$s$^{-1}$. $f_{\rm{sed}}$ is kept constant at 0.5. This shows that higher eddy diffusion coefficients result in the production of larger particles.}
    \label{fig:p_vs_r_fixed_fsed}
\end{figure*}

\begin{figure*}
    \includegraphics[width=\textwidth]{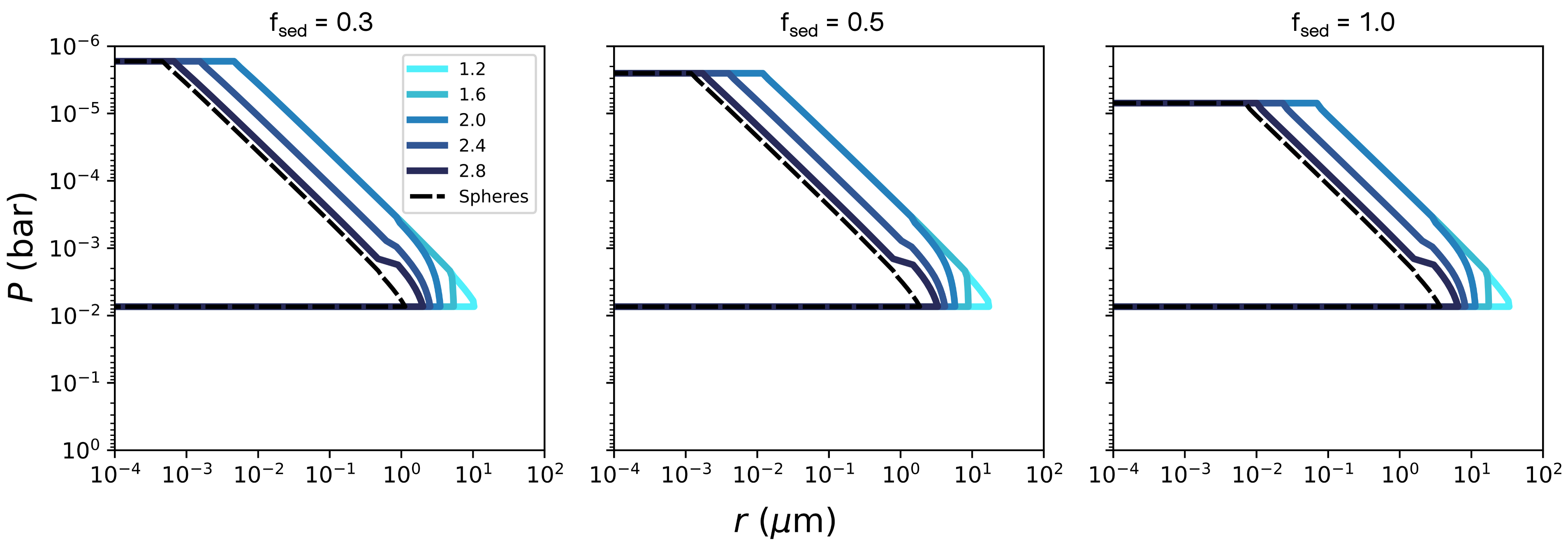}
    \caption{Mean sizes of KCl particles (compact radius) in cloud layers of a warm Neptune atmosphere, with $f_{\rm{sed}}=0.3$, $0.5$ and $1$ (from left to right). $K_{\rm{zz}}$ is kept constant at $10^9$~cm$^2$s$^{-1}$. This shows that higher $f_{\rm{sed}}$ values result in larger particles and thinner clouds that do not extend as high in altitude.}
    \label{fig:p_vs_r_fixed_Kzz}
\end{figure*}

\begin{figure*}
    \includegraphics[width=\textwidth]{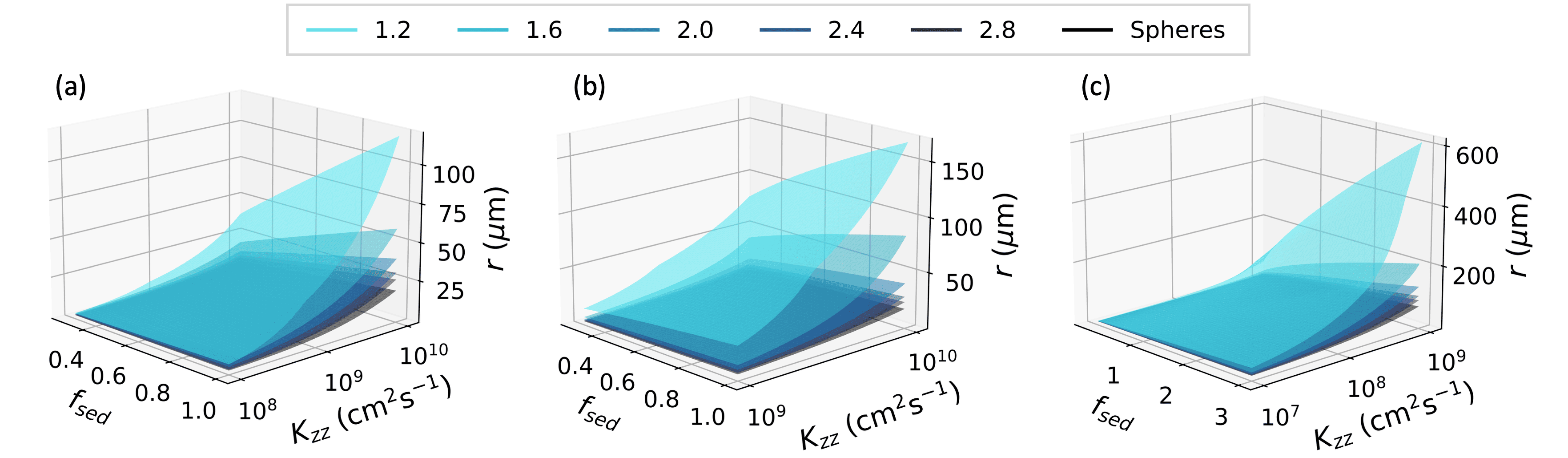}
    \caption{Compact radii of the largest particles as a function of $K_{\rm{zz}}$ and $f_{\rm{sed}}$ for our (a) warm Neptune, (b) hot Jupiter and (c) brown dwarf. In all cases, the lowest fractal dimensions form the most massive particles. Higher values of $K_{\rm{zz}}$ and $f_{\rm{sed}}$ both cause particle size to increase.}
    \label{fig:3D_largest_particles}
\end{figure*}

\subsection{Overall Opacity Changes} 

Figure~\ref{fig:p_vs_r_fixed_fsed} demonstrates how increasing $K_{\rm{zz}}$ increases the mean particle size within each cloud layer. This is intuitive to understand -- increasing the strength of the upwards transport means that particles need to have more mass (and be larger) to have an increased terminal velocity, which can then balance the increased $w_*$. Similarly, Figure~\ref{fig:p_vs_r_fixed_Kzz} demonstrates how increasing $f_{\rm{sed}}$ increases the mean particle size within each cloud layer, which is a result of the simple power law relationship between the particle size and $f_{\rm{sed}}$ (see Eq. 17 of \citetalias{moran_and_lodge_2025} -- increasing $f_{\rm{sed}}$ directly causes bigger particles). To show that this is universal, Figure~\ref{fig:3D_largest_particles} shows the largest particle formed for each shape as a function of both $K_{\rm{zz}}$ and $f_{\rm{sed}}$ in each of the warm Neptune, hot Jupiter and brown dwarf cases. Two key effects can be seen in Figure~\ref{fig:3D_largest_particles}:

\begin{enumerate}
    \item Increasing either of $K_{\rm{zz}}$ and $f_{\rm{sed}}$ causes particle sizes to increase.
    \item When aggregates are modeled with lower fractal dimensions, they form into larger particles.
\end{enumerate}

Point 1 is especially important, because it highlights how $K_{\rm{zz}}$ and $f_{\rm{sed}}$ can control whether the particles form in the small particle regime (section \ref{sec:rules_small_particles}), large particle regime (section \ref{sec:rules_large_particles}) or somewhere in-between.

\begin{figure*}
    \includegraphics[width=\textwidth]{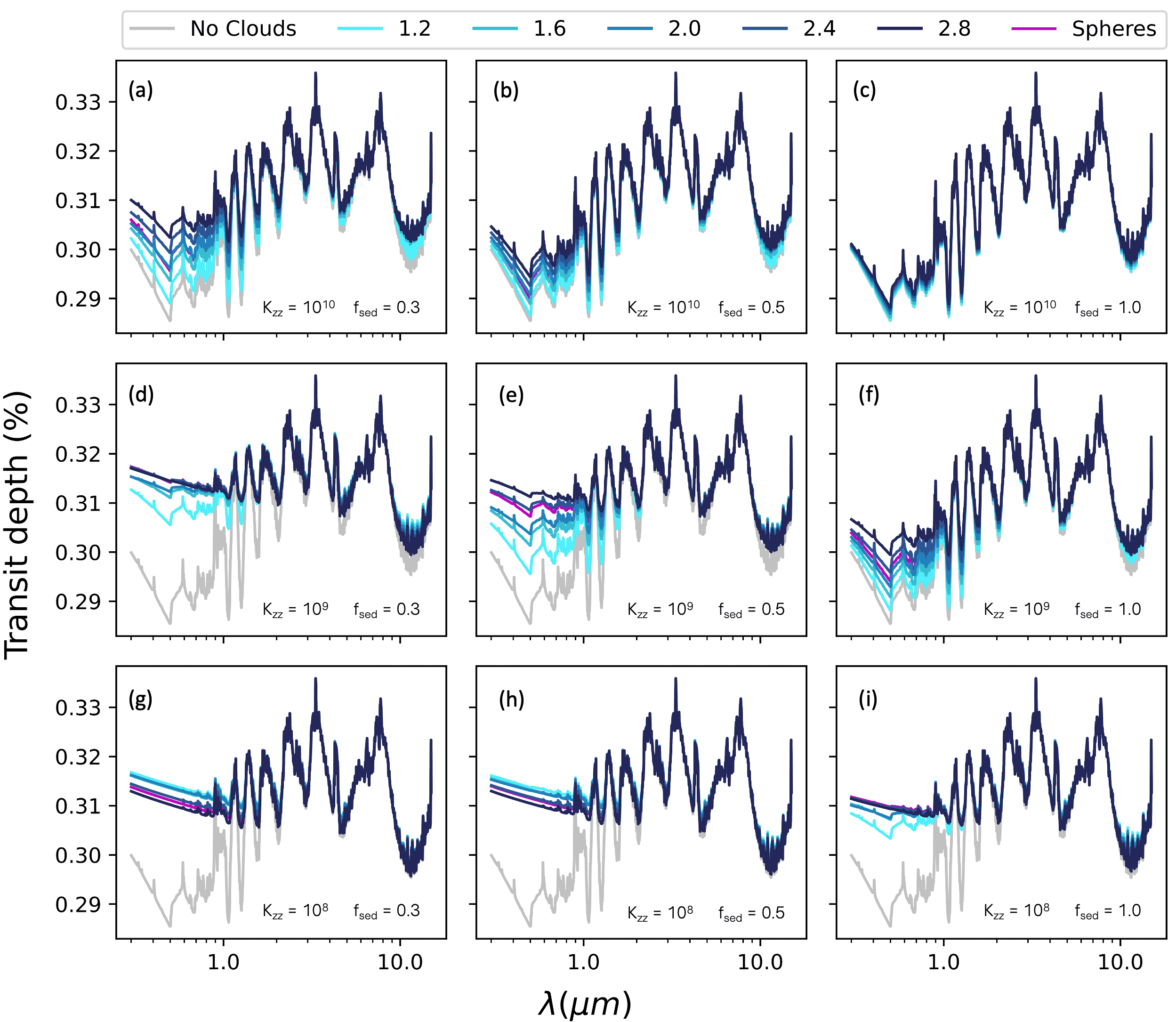}
    \caption{Transmission spectra for KCl clouds in a warm Neptune atmosphere, for a range of $K_{\rm{zz}}$ (cm$^2$s$^{-1}$) and $f_{\rm{sed}}$ values. As $K_{\rm{zz}}$ increases (upwards in columns) or $f_{\rm{sed}}$ increases (to the right in rows), particle sizes become larger. In panels (g) and (h), particles are small (relative to $\lambda$) and so the spectra for $D_{\rm{f}}=1.2$ are more opaque than for $D_{\rm{f}}=2.8$. In all other panels, particle sizes are larger and so spectra for $D_{\rm{f}}=1.2$ are less opaque than for $D_{\rm{f}}=2.8$. An interactive version of this plot is available -- see footnote~\ref{footnote:spectra}.}
    \label{fig:Neptune_9_panel}
\end{figure*}

\begin{figure*}
    \includegraphics[width=\textwidth]{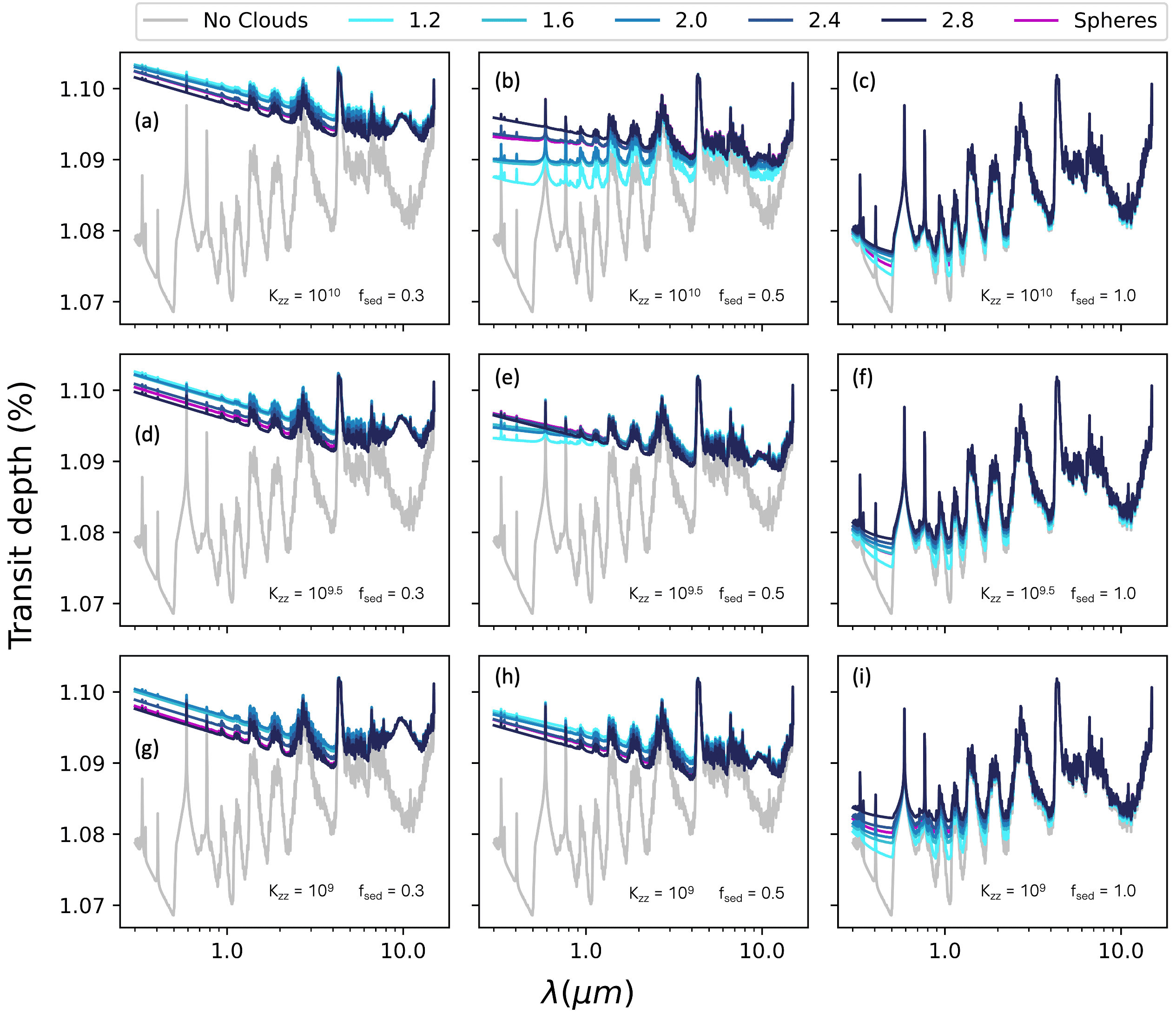}
    \caption{Transmission spectra for Mg$_2$SiO$_4$ clouds in a hot Jupiter atmosphere, for a range of $K_{\rm{zz}}$ (cm$^2$s$^{-1}$) and $f_{\rm{sed}}$ values. As $K_{\rm{zz}}$ increases (upwards in columns) or $f_{\rm{sed}}$ increases (to the right in rows), particle sizes become larger. In panels (a), (d), (g) and (h), particles are small (relative to $\lambda$) and so the spectra for $D_{\rm{f}}=1.2$ are more opaque than for $D_{\rm{f}}=2.8$. In all other panels, particle sizes are larger and so spectra for $D_{\rm{f}}=1.2$ are less opaque than for $D_{\rm{f}}=2.8$. An interactive version of this plot is available -- see footnote~\ref{footnote:spectra}.}
    \label{fig:Jupiter_9_panel}
\end{figure*}

\begin{figure*}
    \includegraphics[width=\textwidth]{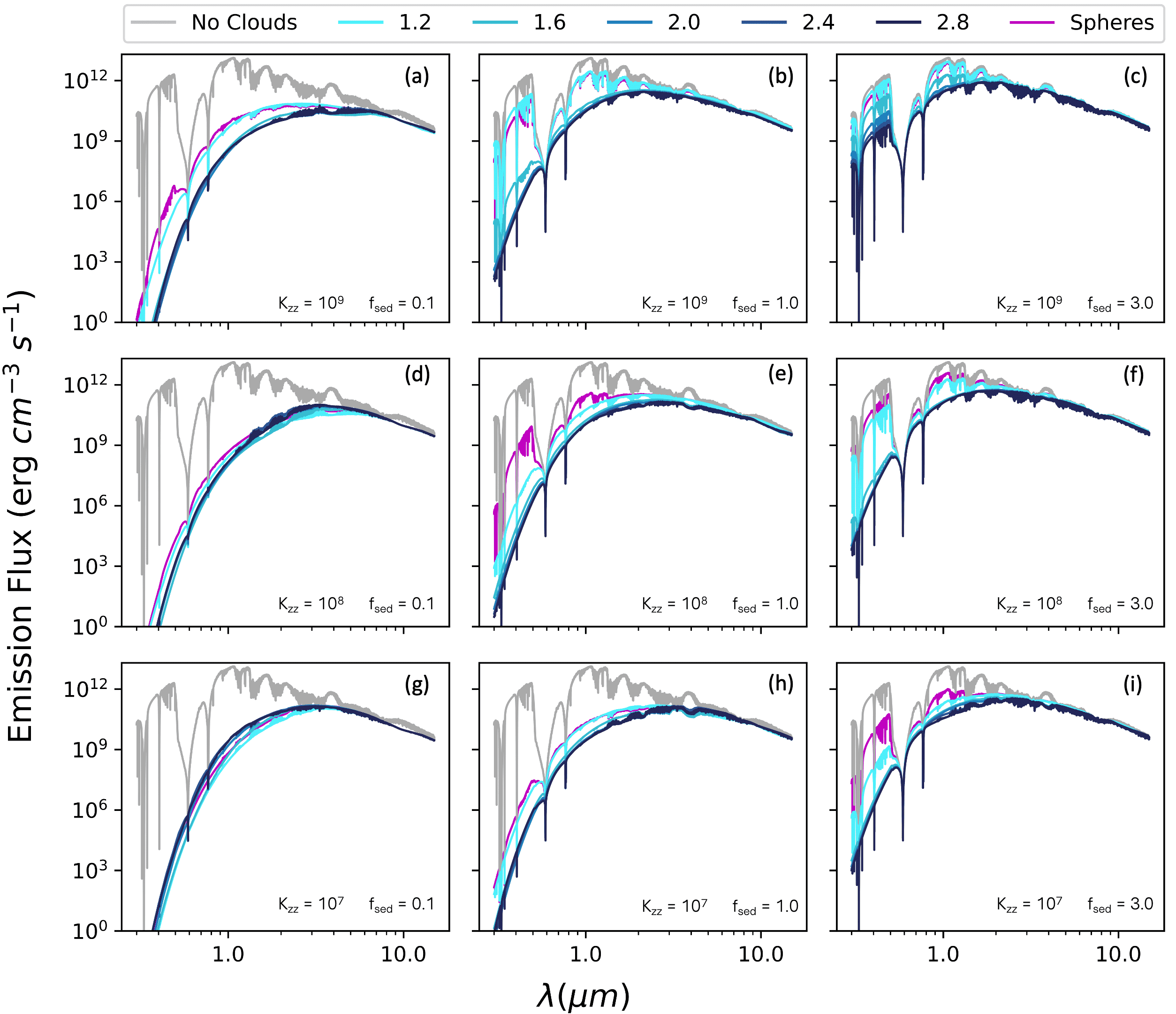}
    \caption{Emission spectra for Mg$_2$SiO$_4$ clouds in a $T_{\rm{eq}}=1700$~K brown dwarf atmosphere, for a range of $K_{\rm{zz}}$ (cm$^2$s$^{-1}$) and $f_{\rm{sed}}$ values. As $K_{\rm{zz}}$ increases (upwards in columns) or $f_{\rm{sed}}$ increases (to the right in rows), particle sizes become larger. In panel (g) and for the region where $\lambda>1~\mu$m in panel (d), particles are small relative to $\lambda$ and so spectra for $D_{\rm{f}}=1.2$ are more opaque than for $D_{\rm{f}}=2.8$, meaning less flux is emitted overall. In all other panels, particle sizes are larger and so spectra for $D_{\rm{f}}=1.2$ are less opaque than for $D_{\rm{f}}=2.8$, resulting in more flux being emitted. An interactive version of this plot is available -- see footnote~\ref{footnote:spectra}.}
    \label{fig:Brown_dwarf_9_panel}
\end{figure*}

The effect of particle regime can be directly observed in the spectra. In Figures~\ref{fig:Neptune_9_panel}(g) and \ref{fig:Neptune_9_panel}(h), for the case of KCl clouds in a warm Neptune, $K_{\rm{zz}}$ and $f_{\rm{sed}}$ is low and so the particles formed are still in the ``small particle'' regime. Therefore, following Eq.~\ref{eq:opacity_rayleigh} ($\kappa \propto r^3$) , the elongated fractals (which form into more massive particles than the compact shapes, and therefore have larger values of compact radius $r$) have the highest opacity, and cause the largest transit depths. However, in all other Figure~\ref{fig:Neptune_9_panel} panels, $K_{\rm{zz}}$ or $f_{\rm{sed}}$ are larger, and so the particles formed are also larger. The particles here are in the large particle regime, and their opacity behaves according to Eq.~\ref{eq:opacity_large_particles} ($\kappa \propto \frac{1}{r}$) -- the elongated fractals are still more massive than the compact shapes (and have larger compact radius $r$), but in this regime the large elongated fractals have the least opacity (and the smallest transit depth). 
Because the behavior inverts between the two particle size regimes, there is a transition between them -- to visualize this more easily, we provide an interactive version of Figures~\ref{fig:Neptune_9_panel}, \ref{fig:Jupiter_9_panel} and \ref{fig:Brown_dwarf_9_panel} where users are able to vary $K_{\rm{zz}}$ and $f_{\rm{sed}}$ gradually to values of their choice\footnote{\url{https://www.star.bris.ac.uk/spectra}\label{footnote:spectra}}. Using these, the transition from the small particle to large particle regime can be seen very clearly.

In our hot Jupiter environment (Fig.~\ref{fig:Jupiter_9_panel}), even though we are considering a different condensing species (Mg$_2$SiO$_4$ clouds) and a very different atmosphere/pressure-temperature profile, the results are very similar. In this case, panels in Figure~\ref{fig:Jupiter_9_panel}(a), (d), (g) and (h) show the effects of the ``small particle'' regime where elongated fractals are the most opaque shapes, whereas panels in Figure~\ref{fig:Jupiter_9_panel}(b), (c), (f) and (i) (where $K_{\rm{zz}}$ and $f_{\rm{sed}}$ are higher) are in the ``large particle'' regime where compact aggregates are the most opaque shapes. Panel ~\ref{fig:Jupiter_9_panel}(e) is particularly demonstrative, because the elongated shapes are the most opaque at short wavelengths but the least opaque at long wavelengths. This occurs because the conditions for whether a particle is in the small or large particle regime depend on wavelength -- and so the smallest wavelengths will enter the ``large particle'' regime first (because a given particle will appear larger relative to smaller wavelengths). Again, this transition between the regimes, including the wavelength dependence, can be clearly and smoothly seen in the interactive versions of these Figures (see footnote~\ref{footnote:spectra}).

Even for the emission spectra of brown dwarf atmospheres containing Mg$_2$SiO$_4$ clouds (Fig.~\ref{fig:Brown_dwarf_9_panel}), the same principle applies. For the small values of $K_{\rm{zz}}$ and $f_{\rm{sed}}$ in panels \ref{fig:Brown_dwarf_9_panel}(d) and \ref{fig:Brown_dwarf_9_panel}(g), the elongated fractals are the most opaque, and therefore they block the most light, causing the emission spectra to be lower in intensity than for the compact aggregate case. In all other panels the particles are larger, and therefore the linear fractals become least opaque, allowing the highest flux intensity through (and so here the emission spectra are highest in value). 

Spherical particles do not necessarily fit the pattern found in the aggregates. As the most compact shape, it may be tempting to assume that spheres should always sit near or beyond the spectra for $d_f=2.8$. However, as seen in Figure~\ref{fig:Q_ext_vs_r}, even the most compact aggregates can have very different optical properties to spheres. To demonstrate why, we combine Eq.~\ref{eq:N_spheres} and Eq.~\ref{eq:kappa} to give:

\begin{equation} \label{eq:kappa_sphere_aggregate_comparison}
    \kappa \propto \frac{Q_{\rm{ext}}}{r}
\end{equation}

Choosing one example to illustrate, the largest spheres in any pressure layer from Figure~\ref{fig:Neptune_9_panel} have mean particle sizes of $r=4.1~\mu$m, with $Q_{\rm{ext}}=2$ at a wavelength of $0.2~\mu$m. In comparison, the most compact aggregates ($D_{\rm{f}}=2.8$) have particle sizes of $r=6.9~\mu$m in this layer, where $Q_{\rm{ext}}=7.9$. Finally, the least compact fractals ($D_{\rm{f}}=1.2$) form particle sizes of $r=33.8~\mu$m in this layer, and their extinction efficiency $Q_{\rm{ext}}=6.5$. Substituting these values into Eq.~\ref{eq:kappa_sphere_aggregate_comparison}, we find that the compact aggregates are 2.3$\times$ more opaque than the spherical particles, and the least compact aggregates are 2.5$\times$ less opaque than the spheres. The $D_{\rm{f}}=2.8$ aggregates have higher opacities because the $Q_{\rm{ext}}$ value for $D_{\rm{f}}=2.8$ is so much higher than for spheres at these particle sizes (see right-side of Fig.~\ref{fig:Q_ext_vs_r}). In this example, spheres behave more similarly to fractal aggregates of $D_{\rm{f}}=2$ in their effect on transmission spectra. In that sense, spectra for aggregate models can exhibit a \textit{range} either side of the spherical model rather than a trend in a single direction away from them as particles get less compact. 

We note that here we are, in effect, exploring the influence of particle porosity on transmission spectra. Studies of opacity changes induced by particle porosity have a long history in the protoplanetary disk and instellar medium literature \citep[e.g.,][]{henning1996,min2007,kataoka2014,min2016multiwavelength}; see also the recent review by \citet{potapov2025}. Our work extends these insights into applications for transiting and directly imaged exoplanets and brown dwarfs.

\subsection{Scattering Specific Changes in Opacity}

\begin{figure}
    \includegraphics[width=\columnwidth]{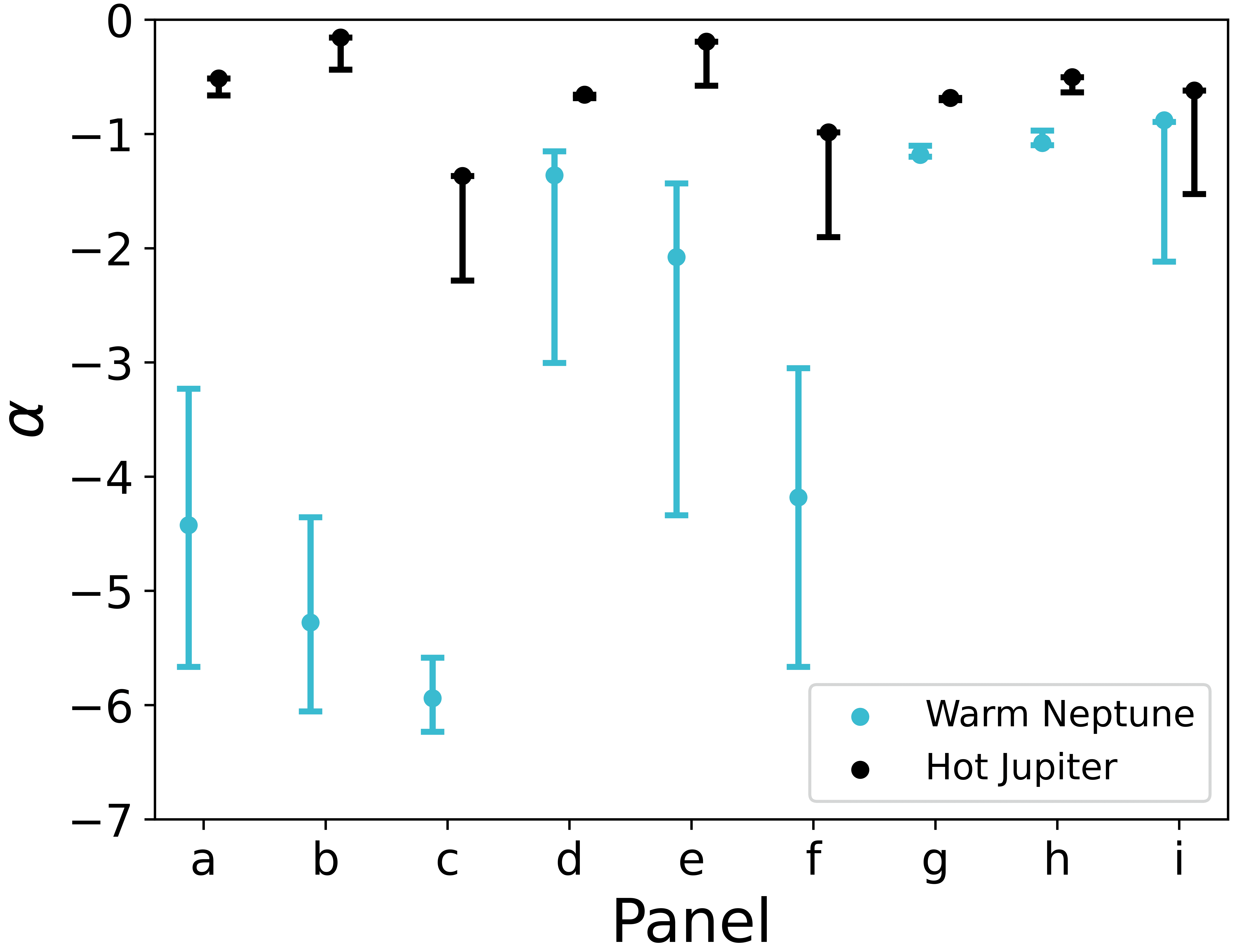}
    \caption{Values of $\alpha$ for each panel in Figures \ref{fig:Neptune_9_panel} and \ref{fig:Jupiter_9_panel}, measured from the gradient between $0.3-0.5~\mu$m. Circular markers represent the value of $\alpha$ for spherical cloud particles, whereas vertical lines mark the range of $\alpha$ for fractal aggregates with $1.2<d_{\rm{f}}<2.8$ for each case of $K_{\rm{zz}}$ and $f_{\rm{sed}}$.}
    \label{fig:gradient_analysis}
\end{figure}

The scattering slope is typically described by the parameter $\alpha$, which is defined \citep{des2008rayleigh} by:
\begin{equation} \label{eq:scattering_slope}
	\alpha = H^{-1}\frac{\rm{d}R_{p}}{\rm{d}\ln{\lambda}},
\end{equation}
where $R_{p}$ is the radius of the planet at wavelength $\lambda$, and $H$ is the scale height. The scale height $H$ varies with altitude throughout our model atmospheres, however it becomes roughly constant higher in the atmosphere where the clouds form and persist. For the warm Neptune case, we use $H=2.78\times 10^5$~m, which is the average value where $P<10^{-2}$~bar. For the hot Jupiter case we use $1.59\times 10^5$~m, which is the average scale height between $10^{-3}\leq P \leq 10^{-1}$~bar. An atmosphere dominated by purely-scattering spherical particles would have $\alpha=-4$, and purely absorbing spherical particles would have $\alpha=-1$. However, real atmospheres are much more complex, and scattering slopes between 0 and $-35$ have been measured (see \citealt{ohno2020super} for a discussion of these). To quantitatively demonstrate the difference in scattering slope between models in Figures~\ref{fig:Neptune_9_panel} and \ref{fig:Jupiter_9_panel}, Figure~\ref{fig:gradient_analysis} shows the value of $\alpha$ obtained for spherical particles in each case of $K_{\rm{zz}}$ and $f_{\rm{sed}}$. The maximum and minimum value of $\alpha$ for fractal aggregates are also plotted as vertical lines. Fractal aggregates create a range of scattering slopes either side of the spherical value, further emphasizing the need to allow fractal dimension $d_{\rm{f}}$ to vary as a free parameter in retrievals and forward models when interpreting observational spectra. The scattering slope $\alpha$ is also dependent on the cloud structure, in particular the cloud scale height with respect to the gas phase atmospheric scale height, with fractal aggregation another nuance to consider in the derivation of the scattering slope.

While the above plots can be instructive and demonstrative of the general rules, we urge readers to interpret the extreme cases with caution. The largest $K_{\rm{zz}}$ and $f_{\rm{sed}}$ would require very large particles to balance the upwards forces and exist as persistent particles within pressure layers in the atmosphere. The largest particles for $D_{\rm{f}}=1.2$ in panel Fig \ref{fig:Brown_dwarf_9_panel}c would be $r>600~\mu$m, which is almost a single long chain $>6$~cm across. In reality, these particles would almost certainly compress as they grow, especially for the lowest fractal dimensions (\citealt{ohno2020clouds}; \citetalias{moran_and_lodge_2025}), and thus these results for this extreme case are probably not realistic. However, we present them here to demonstrate the boundaries of the model, and simply encourage users to use the diagnostic output tools built into {\virga} to analyze their results in detail, and to take care when interpreting results for individual cases.

\section{Conclusions} \label{sec:Conclusions}

We find that fractal aggregates have significant and measurable differences in transmission and emission spectra, when compared to models that make the assumption that all particles are spherical. These differences arise as a direct consequence of the following:

\begin{enumerate}
    \item Aggregates form into larger particles than spheres, because their terminal velocity is smaller for a given mass (and therefore they need to have more mass for their downwards velocity to equal a given upwards velocity within a pressure layer).
    \item Aggregates have very different optical properties to spherical particles.
\end{enumerate}

\noindent Points 1 and 2 combine to give transmission and emission spectra that are very different compared to when particles are modeled as spheres. We find detectable, measurable differences in spectra (even for the most compact and almost spherical fractal aggregates), and changes in the scattering slope of factors up to $\approx2$. In addition, we find that there are two broad rules for purely scattering (non-absorbing) chemical species:

\begin{enumerate}[(i)]
    \item If aggregates are small (relative to the wavelength of light), elongated aggregates will be more opaque than compact aggregates (section \ref{sec:rules_small_particles}).
    \item If aggregates are large (relative to the wavelength of light), elongated aggregates will be less opaque than compact aggregates (section \ref{sec:rules_large_particles}).
\end{enumerate}

These rules hold for KCl aggregates in typical warm Neptunes, Mg$_2$SiO$_4$ in typical hot Jupiters, and Mg$_2$SiO$_4$ in typical L-type brown dwarf atmospheres. These rules are designed to help the exoplanet community interpret observations by providing intuition about how particle shape might influence spectra. We strongly encourage the community to fully explore the new parameter $D_{\rm{f}}$ in {\virga}, which has been shown here to significantly affect spectra, and in many circumstances will help determine a greatly improved match between theory and real exoplanet/brown dwarf data. We aim to analyze specific measured exoplanet spectra in more depth in a future paper, and encourage all future case-studies to consider the role of particle shape on their spectra.

\begin{acknowledgments}
We sincerely thank an anonymous reviewer for their thorough and thoughtful feedback, which appreciably improved this manuscript. M.G.L acknowledges the generous support of the University of Bristol Keith Burgess Scholarship and Frederick Frank Fund. 
S.E.M. is supported by NASA through the NASA Hubble Fellowship grant HST-HF2-51563 awarded by the Space Telescope Science Institute, which is operated by the Association of Universities for Research in Astronomy, Inc., for NASA, under contract NAS5-26555. 
ZML acknowledge the financial support from the Science and Technologies Facilities Council grant number ST/V000454/1.
H.R.W. was funded by UKRI under the UK government’s Horizon Europe funding guarantee for an ERC StG [grant number EP/Y006313/1]. This work benefited from in-person collaboration between M.G.L and S.E.M funded by the the same ERC StG [grant number EP/Y006313/1].

\end{acknowledgments}

\software{\texttt{virga} \citep{batalha2025,Batalha2020, Rooney2022}, \texttt{PICASO} \citep{picaso,Mukherjee2023}, \texttt{NumPy} \citep{harris2020array}, \texttt{matplotlib} \citep{Hunter:2007}, \texttt{SciPy} \citep{2020SciPy-NMeth}}



\bibliography{references}{}
\bibliographystyle{aasjournalv7}



\end{document}